\documentstyle[preprint,aps,prl]{revtex}

\begin{document}
\draft

\title{First Order Phase Transitions in Gravitational Collapse}

\author{Piotr Bizo\'n${*}$}
\address{Department of Mathematics, University of Michigan, Ann Arbor,
 Michigan}
\author{Tadeusz Chmaj}
\address{Institute of Nuclear Physics, Cracow, Poland}

\date{\today}

\maketitle

\begin{abstract}
In recent numerical simulations of spherically symmetric gravitational collapse
a new type of critical behaviour, dominated by a sphaleron
solution, has been found. In contrast to the previously studied models, in this
case there is a finite gap in the spectrum of black-hole masses which is
reminiscent of a first order phase transition. We briefly summarize the
essential features of this phase transition and describe the basic heuristic
picture underlying the numerical phenomenology.

\end{abstract}
\pacs{04.25.Dm, 04.40.-b, 04.70.Bw}


One of the main open problems in classical general relativity is the issue
of global dynamics of solutions of Einstein's equations. The presently
achievable mathematical techniques appear to be insufficient to address this
problem in full generality, so most researchers have focused their attention on
a more tractable case of spherical symmetry. In particular,
in a remarkable series of papers (see~\cite{c1}, \cite{c2} and references
therein) Christodoulou 
has analysed
the evolution of regular initial data for the spherically symmetric
Einstein-Klein-Gordon equations. He showed that for "weak" initial data
there exists a unique global solution which asymptotes to the Minkowski
spacetime, whereas "strong" initial data form a singularity, which, in accord
with weak cosmic censorship hypothesis, is surrounded by an event horizon
(here "weak" and "strong" have well defined meaning in terms of a certain
function norm). These results suggested that there is a "critical
surface" in the phase space which separates the two kinds of initial data.
The initial data lying on this critical surface are at the treshold of black
hole formation.
A natural question is: what is the mass of a black hole at the treshold?
Does it continuously decrease to zero, or is there a finite lower bound
(mass gap)? These two possibilities will be referred to as second and first
order phase transitions, respectively.

The pioneering numerical investigations of this problem were carried out
by Choptuik who analysed the evolution of one-parameter
families of initial data crossing the critical surface~\cite{matt}. For each 
family,
labelled by a parameter $p$, Choptuik found a critical value $p^{\star}$
such that the data with $p>p^{\star}$ form a black hole, while the data with
$p<p^{\star}$ do not. It turned out that the marginally supercritical data form
black holes  with masses satisfying the power law $M_{BH} \simeq C
(p-p^{\star})^{\gamma}$ with a universal (i.e. family independent) critical 
exponent $\gamma$.
Therefore, as $p \rightarrow p^{\star}$, the black hole mass decreases 
continuously to zero
which is reminiscent of the second order phase transition. Morever, 
Choptuik discovered that in the intermediate asymptotics (i.e. before a 
solution "decides" whether to collapse or not) all near-critical solutions
approach a universal attractor. This attractor, called the critical solution,
has an unusual symmetry of discrete self-similarity.

Quickly after Choptuik's discovery similar  critical effects have been observed
in other models of gravitational collapse~\cite{ae,ec,hhs}. In all cases the overall picture
of criticality was qualitatively the same as in the scalar field collapse,
possibly with one difference: in certain models the critical solution was
continuously (rather than discretly) self-similar. These studies provided
convincing evidence that such features as universality, black-hole mass
scaling, and self-similarity (discrete or continuous) are the robust
properties of second order phase transitions in gravitational collapse.
Although none of these properties have been proven rigorously, a substantial
progress has been made on a heuristic level. In particular, at present we have
a convincing picture of the origin of universality and scaling
(see~\cite{carsten} for a recent review).

None of the first studied models had a mass/length scale, in analogy
to the vacuum Einstein's equations. This raises the question: which features
of the critical colapse are inherently related to the scale invariance,
or, putting this differently, how does the presence of length/mass scale 
affect the scenario of critical behaviour?  It follows from dimensional
analysis that, under the assumption of universality, the lack of scale
implies that the mass gap must be zero. However, the converse is not true,
as was already noticed by Choptuik in his studies of self-interacting
scalar field. It seems that what matters in the evolution is not the scale 
itself but rather the presence of nonsingular stationary solutions (which are
of course excluded for scale invariant equations).

In order to understand better the role of scale and stationary solutions
in the dynamics of Einstein's equations, we have recently investigated two
models: Einstein-Yang-Mills (in collaboration with Matt
Choptuik~\cite{eym}) and Einstein-Skyrme~\cite{bc2}. Both
these models
possess a mass/length scale (actually the ES model has two scales) and 
static regular solutions.

To make this paper self-contained, we first present the general 
setting for the spherically symmetric evolution.
Studying a spherically symmetric Einstein-matter system it is convenient
to use the following ansatz
for the metric  
\begin{eqnarray}
ds^2 &=& -e^{-2\delta} N dt^2 + N^{-1} dr^2 + r^2 d \Omega^2,
\label{METRIC}
\end{eqnarray}
where $d \Omega^2$ is the standard metric on the unit 2-sphere and $\delta, N$
are functions of $(t,r)$. Although this coordinate system cannot penetrate
an event horizon, 
this is not a serious disadvantage, as Choptuik emphasized, in studying the 
{\em formation} of horizons.
The main advantage of the choice (1) is that the Einstein's equations simplify
considerably in terms of the metric functions $N$ and $\delta$
\begin{eqnarray}
N' &=& \frac{1-N}{r} - 8 \pi G r\,T_{00},
\\
{\dot N} &=& -8 \pi G r e^{-\delta} N \,T_{01},
\\
{\delta'} &=& -4 \pi G r N^{-1} (T_{00}+T_{11}),
\end{eqnarray}
where overdots and primes denote $\partial / \partial t$ and 
$\partial / \partial r$, respectively, and the components of the
stress-energy tensor of matter $T_{ab}$ are expressed in the orthonormal 
frame determined by the
metric (1) ($e_0=e^{\delta} N^{-1/2} \partial_t$, $e_1=N^{1/2} \partial_r$).

The full system of evolution equations consists of Eqs.(2-4) and evolution
equations
for a matter field~\footnote{As it is well known, in many cases the
evolution equations for matter are implicitly contained in the Einstein
equations. In particular, for the two models discussed below the consistency
condition for the Eqs.(2) and (3) is equivalent to the wave equation for a
matter field. }. For the sake of simplicity let us assume that matter is described by
one scalar function $F(r,t)$ which satisfies a nonlinear wave equation (this is
a typical situation).
As a consequence of Birkhoff's theorem the essential dynamics of the system
resides in $F(r,t)$. In order to solve the evolution equations we need to
supplement them with boundary and initial conditions. To ensure regularity
we require that the components of the stress-energy tensor are bounded for
all $(t,r)$ and that $N(0,t)=1+O(r^2)$ at the center. Asymptotic flatness
requires that $N(r,t)=1+O(1/r^2)$ for $r \rightarrow \infty$. The time coordinate is
normalized by the boundary condition $\delta(\infty,t)=0$ (so $t$ is the proper
time at spatial infinity). The initial value problem for the above equations is
solved as follows. At $t=0$ one takes asymptotically flat regular initial data
for the matter field $F(r,0)$ and $\dot F(r,0)$. Then the elliptic
Eqs.(2) and (4) are solved yielding initial metric functions $N(r,0)$ and $\delta(r,0)$.
Once a full set of initial data is constructed, it is evolved  using Eq.(3) and the hyperbolic evolution
equation for $F$. The function $\delta$ is updated at each subsequent
moment of time using Eq.(4). This scheme is called  the free evolution, as opposed
to the alternative scheme of fully constrained evolution in which the hamiltonian
constraint (2), rather than Eq.(3), is used to compute the function $N$.

We are now ready to discuss critical phenomena in the
evolution. Here we focus our attention exclusively on the first order phase 
transitions.
 In passing, we remark that, from
the theoretical perspective, second order phase transitions are much more
interesting
because of their bearing on the cosmic censorship hypothesis. On the other
hand,
first order phase transitions might be more important in the astrophysical
 context.

To make this presentation consize we first describe the basic scenario in 
a model-independent manner and then illustrate it with  concrete models.
For a first order transition to occur in a spherically-symmetric
Einstein-matter system, we need to make 
two assumptions:~\footnote{The importance and consequences of the analogous 
assumptions in the context of second
order phase transitions in critical collapse were first spelt out 
by Koike,
Hara, and Adachi~\cite{kha}.}

{\em Assumption 1.} There exists a static regular
asymptotically flat solution with {\em one} linearly unstable eigenmode (in 
field theory
such solution is referred to as a sphaleron). Let us denote this solution by $X^u$ with
$X$ standing for $(F,N,\delta)$. The uniqueness of the unstable eigenmode 
means that the linear evolution of an initially small spherical perturbation about
$X^u$ can be decomposed into the sum
\begin{equation}
\delta X^u(r,t) = C e^{\lambda t} 
\xi_{\lambda}(r)
+ \sum_{i=1}^{\infty} C_i e^{\mu_i t} \xi_{\mu_i}(r)
\end{equation}
of  the single growing mode with a positive eigenvalue $\lambda$ and
the decaying modes with $Re(\mu_i)<0$. Physically the damping of 
non-growing modes is due to the loss of energy by radiation. Mathematically
this is reflected in  non-self-adjointness of the eigenvalue problem.

{\em Assumption 2.} The ultimate fate of the perturbation (5) depends only
on the sign of the amplitude $C$. For one sign of $C$, say
$C>0$, a black hole forms, while for $C<0$ there exists a global
in time regular evolution. In the latter case the final state depends on the
details of a model -- it might be the Minkowski spacetime (this case is usually
referred to as dispersion), or some stable regular solution. 

Assumption 1 means that that the stable manifold $W_S$ of the solution
$X^u$ has codimension one. Assumption 2 means that $W_S$ is a "critical"
surface in the sense of dividing (locally) the phase space into collapsing and
non-collapsing initial data. 

Now, consider a one-parameter family of initial data $\Phi(r,p)$, where $p$ is
a parameter,
which intersects $W_S$ at some parameter value $p=p^{\star}$. Let $X_p(r,t)$ 
denote the solution corresponding to these initial data. 
The critical initial data are attracted along $W_S$ towards
the solution $X^u$. A near-critical solution, by continuity, remains close to
$W_S$ and, once it  gets close to $X^u$, 
can be approximated by the linearization (5) around $X^u$ with the amplitudes
$C$ and $C_i$ depending on the initial data. Since by definition
$C(p^{\star})=0$, we have
\begin{equation}
X_p(r,t) = X^u(r) + A (p-p^{\star})\; e^{\lambda t} \xi_{\lambda}(r)
+ {\mbox decaying} \quad {\mbox modes},
\end{equation}
where $A=\frac{dC}{dp}(p^{\star})$.
The range of $t$  for which this linear
approximation is valid is called the intermediate asymptotics. In this
asymptotics, the solution $X_p(r,t)$ initially approaches $X^u$ but later the
growing  mode becomes dominant and the solution is repelled from $W_S$ along the
unstable manifold of $X^u$.
Because of this behaviour the solution $X^u$ is sometimes called the intermediate
attractor. 
The duration of the intermediate asymptotics is determined by the time $T$ in
which the unstable mode grows to a finite size
$|p-p^{\star}| e^{\lambda T} \sim O(1)$, which gives $T \sim 
-\lambda^{-1} \ln|p-p^{\star}|$. Thus, the larger $\lambda$, the better fine-tuning is required to see the
solution $X^u$ clearly pronounced as the intermediate attractor.

The scenario of critical collapse  summarized above naturally explains the universality
(that is family-independence) of this phenomenon -- it simply follows from the
fact that the evolution of 
{\em all} near-critical data
is governed by the {\em same} unstable mode around the intermediate attractor.
Within this framework we also see why it is essential that the solution $X^u$
has exactly one unstable mode.
If it was linearly stable (that is if it had no unstable modes),
then it would be an attractor of an open set of initial data and it would not be
related to any critical behaviour. On the other hand if $X^u$ had
two or more unstable modes then a generic one-parameter family of initial
data would not intersect its stable manifold and {\em eo ipso} the 
critical behaviour would not be  generic.

By Assumption 2 all solutions starting with initial data $X^u(r)+\epsilon 
\xi_{\lambda}(r)$ with some small positive amplitude $\epsilon$ form black holes.
As follows from (6), for any $\epsilon$,  such initial data can be extracted from 
the evolution of supercritical solutions with sufficiently small $p-p^{\star}$
at some time $t_p$ satisfying $A (p-p^{\star}) e^{\lambda t_p} = \epsilon$.
Although the time $t_p$ depends on $p$, the evolution for $t>t_p$ is independent
of $p$~\cite{carsten}. Denoting  the mass of a resulting black hole by $m_{BH}(\epsilon)$,
we can define the mass gap as
$m^{\star}=\lim_{\epsilon \rightarrow 0} m_{BH}(\epsilon)$. The mass gap
$m^{\star}$ is bounded from above by the mass $m_u$ of the solution $X^u$.
The difference $m_u-m^{\star}$ can be interpreted as the total energy radiated
away to infinity during the critical collapse.

Now we substantiate the general picture presented above with two models in which
the first order phase transitions were observed:
Einstein-Yang-Mills (EYM) and Einstein-Skyrme (ES).

EYM: We assume the following ansatz for the SU(2)-YM field
\begin{equation}
eF = dw \wedge (\tau_1 d\vartheta + \tau_2 \sin \vartheta d\varphi)
-(1-w^2) \tau_3 d\vartheta \wedge \sin \vartheta d\varphi,
\end{equation}
where $e$ is the coupling constant, $\tau_i$ are the Pauli matrices, and
the YM potential $w$ is a function of $(r,t)$. The evolution
equation for $w(r,t)$ is
\begin{equation}
-(e^{\delta} N^{-1} \dot w)^{\dot{} } + (e^{-\delta} N w')'
+\frac{1}{r^2} e^{-\delta} w (1-w^2) = 0,
\end{equation}
while the Einstein equations have the form (2)-(4) with the stress-energy
tensor
\begin{equation}
T_{00} = \frac{1}{4 \pi e^2 r^2}
\left(N w'{}^2 + e^{2 \delta} N^{-1} \dot w^2 + \frac{(1-w^2)^2}{2
r^2}\right),
\end{equation}
\begin{equation}
T_{11} = \frac{1}{4 \pi e^2r^2}
\left(N w'{}^2 + e^{2 \delta} N^{-1} \dot w^2 - \frac{(1-w^2)^2}{2
r^2}\right),
\end{equation}
\begin{equation}
T_{01} = \frac{1}{2 \pi e^2r^2} e^{\delta}
\dot w w'.
\end{equation}
The EYM equations have a countable family of static asymptotically flat
regular solutions $X_n$ ($n \in N$) discovered numerically by Bartnik and
McKinnon~\cite{bartnik} and later proven rigorously to exist by Smoller
and Wassermann~\cite{sw}. Within the ansatz (7) the solution $X_n$ has $n$ 
unstable
modes, hence the first Bartnik-McKinnon solution $X_1$ satisfies the 
Assumption~1~\cite{sz}. Morever, the nonlinear instability analysis of this solution
performed by Zhou and Straumann~\cite{zs} strongly suggested that the
Assumption~2 is also true. In fact, Choptuik and the present authors showed
 that
for some initial data the solution $X_1$ acts as the intermediate
attractor and controls the first order phase transition~\cite{eym}.
Since vacuum is the only stable solution, the subcritical solutions
disperse.
For supercritical solutions the mass gap was found to be equal (up to 0.1\%) to the mass of $X_1$
(the mass scale is given by $1/(e \sqrt{G})$).

ES:  In this model matter is described an
$SU(2)$-valued scalar function $U$ (called a chiral field).
In spherical symmetry $U=exp(i\vec\tau \cdot
\hat r F(r,t))$
with the dynamics of $F(r,t)$  governed by the equation
\begin{equation}
-(u e^{\delta} N^{-1} \dot F)^{\dot{} } + (u e^{-\delta} N F')'
= \sin(2F) e^{-\delta} \left(
f^2+ \frac{1}{e^2} (\frac{\sin^2{F}}{e^2 r^2}+ N F'^2-
e^{ 2\delta} N^{-1} \dot F^2)\right),
\end{equation}
where $u=f^2 r^2 + 2 \sin^2{F}/e^2$, and $f$ and $e$ are coupling
constants.
The components of stress-energy tensor in the Einstein equations (2)-(4) are
\begin{equation}
T_{00} = \frac{u}{2 r^2} (N F'^2+N^{-1} e^{-2\delta} \dot F^2)
+ \frac{\sin^2{F}}{r^2} (f^2+\frac{\sin^2{F}}{2 e^2 r^2}),
\end{equation}
\begin{equation}
T_{11} = \frac{u}{2 r^2} (N F'^2+N^{-1} e^{-2\delta} \dot F^2)
- \frac{\sin^2{F}}{r^2} (f^2+\frac{\sin^2{F}}{2 e^2 r^2}),
\end{equation}
\begin{equation}
T_{01} = \frac{u}{r^2} e^{\delta} \dot F F',
\end{equation}
Regularity at the center requires that $F(0,t)=0$, while asymptotic
flatness requires that $F(\infty,t)=B \pi$, where an integer $B$, called
the baryon number, is equal to the topological degree of the chiral field.
As long as no horizon forms, the baryon number
is conserved during the evolution, so we have topological selection rules
for the possible end states of given initial data.  The number of
static regular solutions of the ES equations depends on the dimensionless 
parameter
$\alpha=4\pi G f^2$ (this parameter is the square of the ratio
of two length scales $\sqrt{G}/e$ and $1/\sqrt{4 \pi} e f$). For large
values of $\alpha$ there are no regular static solutions. As $\alpha$
decreases, in each
topological sector there is a countable sequence of bifurcations at which 
there appear pairs
of static regular solutions~\cite{bc1}. Here we briefly summarize our results
in
the topological sectors of baryon number zero and one (see~\cite{bc2} for the
details).

In the $B=0$ sector for 
$\alpha<\alpha_0 \simeq  0.00147$ there exists a static regular solution
satisfying the Assumptions~1 and 2 which plays the role of an intermediate
attractor in the critical collapse of specially prepared initial
data. Since the vacuum ({\em i.e.} the Minkowski spacetime) is 
the only regular stable $B=0$ solution, the
subcritical solutions disperse, as in the EYM case.
The case $B=1$ is more interesting.  
Here, for $\alpha<\alpha_1 \simeq 0.040378$, there is a pair of
regular static solutions $X^s$ and $X^u$. The solution $X^s$ is linearly
stable while $X^u$ has one unstable mode~\cite{hds,bc1}. In the limit 
$\alpha \rightarrow 0$ the solution $X^s$ tends to the flat space
skyrmion while $X^u$ has no regular limit. Again, the solution
$X^u$ satisfies the Assumption~2 and plays the role of an intermediate
attractor. However, now the dispersion is topologically forbidden, and
instead the subcritical solutions decay into the stable solution $X^s$ (this was
observed previously in~\cite{hsz}). The solution $X^u$ has larger
mass than  $X^s$ so during its
decay the excess energy has to be  radiated away to
infinity. The  process of settling down to $X^s$ has the form of damped
oscillations (quasinormal ringing). The relaxation time increases
with $\alpha$ and tends to infinity as $\alpha \rightarrow
\alpha_1$, where the solutions $X^s$ and $X^u$ coalesce.
The evolution of supercitical solutions also depends on the baryon number.
For $B=1$ almost no energy is lost during the critical collapse, hence the
mass gap is equal to the mass of $X^u$, in analogy to the EYM case.
In contrast, for $B=0$ supercritical data 
a substantial amount of energy is radiated away, and consequently
the mass gap is smaller than the mass of the unstable solution.
For example, $m^{\star} \simeq 0.76 m_u$ when  $\alpha=0.00145$.

In conclusion, the numerical results in the EYM and ES models
(and also the results of~\cite{brady}) give strong evidence that our
 understanding of first order
phase transitions in gravitational collapse based on the Assumptions~1 and 2 
is correct. In particular, in both cases the formula (6) was shown to
approximate very well the evolution in the intermediate asymptotics.
Actually, this formula was used to reproduce with good accuracy the results
of linear stability analysis. Let us close with the discouraging remark that a 
rigorous
description of the phenomenon described in this paper does not seem feasible
to us, because it would require to overcome a number of mathematical problems
which have not been solved even for  much simpler systems.

{\em Acknowledgments.} 
This research was supported in part by the KBN grant
PB 0099/PO3/96/11. 


\end{document}